\documentclass[runningheads,a4paper]{llncs}

\usepackage[utf8]{inputenc}
\usepackage{amsmath}
\usepackage{amssymb}

\usepackage{graphicx}
\usepackage{url}

\usepackage{tikz}
\usetikzlibrary{matrix,backgrounds}

\DeclareMathOperator{\LCP}{\textsc{Lcp}}

\usepackage{todonotes}

\usepackage{thmtools}

\title{Compressed Communication Complexity of  Longest Common Prefixes}

\titlerunning{Compressed Communication Complexity of Longest Common Prefixes}

\author{Philip Bille\inst{1}\and Mikko Berggreen Ettienne\inst{1}\and Roberto Grossi\inst{2} \and Inge Li Gørtz\inst{1}\and Eva Rotenberg\inst{1}}

\authorrunning{P. Bille\and M. B. Ettienne\and R. Grossi \and I. L. Gørtz \and E. Rotenberg\orcidID{0000-0001-5853-7909}}

\institute{DTU Compute, AlgoLoG, Technical University of Denmark,
   2800 Kgs. Lyngby, Denmark \and Dipartimento di Informatica, Università di Pisa, Italy}

\begin{document}

\maketitle

\begin{abstract}
We consider the communication complexity of fundamental longest common prefix $(\LCP)$ problems. In the simplest version, two parties, Alice and Bob, each hold a string, $A$ and $B$, and we want to determine the length of their longest common prefix $\ell=\LCP(A,B)$ using as few rounds and bits of communication as possible. We show that if the longest common prefix of $A$ and $B$ is compressible, then we can significantly reduce the number of rounds compared to the optimal uncompressed protocol, while achieving the same (or fewer) bits of communication. 
Namely, if the longest common prefix has an LZ77 parse of $z$ phrases, only $O(\lg z)$ rounds and $O(\lg \ell)$ total communication is necessary.
We extend the result to the natural case when Bob holds a set of strings $B_1, \ldots, B_k$, and the goal is to find the length of the maximal longest prefix shared by $A$ and any of $B_1, \ldots, B_k$. Here, we give a protocol with $O(\log z)$ rounds and $O(\lg z \lg k + \lg \ell)$ total communication. 
We present our result in the public-coin model of computation but by a standard technique our results generalize to the private-coin model. Furthermore, if we view the input strings as integers the problems are the greater-than problem and the predecessor problem. 
\end{abstract}


\section{Introduction}
Communication complexity is a basic, useful model, introduced by Yao~\cite{Yao:1979}, which quantifies the total number of bits of communication and rounds of communication required between two or more players to compute a function, where each player holds only part the function's input. A detailed description of the model can be found, for example, in the book by Kushilevitz and Nisam~\cite{KushilevitzNisan1997}. 

Communication complexity is widely studied and has found application in many areas, including problems such as equality, membership, greater-than, and predecessor (see the recent book by Rao and Yehudayoff~\cite{RaoYehudayoff18}). For the approximate string matching problem, the paper by Starikovskaya~\cite{Starikovskaya17} studies its deterministic one-way communication complexity, with application to streaming algorithms, and provides the first sublinear-space algorithm. Apart from these results, little work seems to have been done in general for the communication complexity of string problems~\cite{Starikovskaya18}.


In this paper, we study the fundamental \emph{longest common prefix problem}, denoted \textsc{Lcp}, where Alice and Bob each hold a string, $A$ and $B$ and want to determine the length of the longest common prefix of $A$ and $B$, that is, the maximum $\ell \geq \ell$, such that $A[1..\ell] = B[1..\ell]$ (where $\ell = 0$ indicates the empty prefix). This problem is also called the \emph{greater than problem}, since if we view both $A$ and $B$ as integers, the position immediately after their longest common prefix determines which is larger and smaller. The complexity is measured using the number of rounds required and the total amount of bits exchanged in the communication. An optimal randomized protocol for this problem uses $O(\lg n)$ communication and $O(\lg n)$ rounds~\cite{Smirnov1988,Nisan1993} where $n$ is the length of the strings. Other trade-offs between communication and rounds are also possible~\cite{senaVenkateshb2008}. Buhrman et al.~\cite{Buhrman} describe how to compute \textsc{Lcp} in $O(1)$ rounds and $O(n^\epsilon)$ communication.

We show that if $A$ and $B$ are compressible we can significantly reduce the number of needed rounds while simultaneously matching the $O(\lg n)$ bound on the number of bits of communication. With the classic and widely used Lempel-Ziv 77 (LZ77) compression scheme~\cite{Ziv1977} we obtain the following bound.  

\begin{restatable}{theorem}{lcp}
    \label{thm:lcp}
    The \textsc{Lcp} problem has a randomized public-coin $O(\lg z)$-round protocol with $O(\lg \ell)$ communication complexity, where $\ell \leq n$ is the length of the longest common prefix of $A$ and $B$ and $z \leq \ell$ is the number of phrases in the LZ77 parse of this prefix.
\end{restatable}

Compared to the optimal uncompressed bound we reduce the number of rounds from $O(\lg n)$ to $O(\lg z)$ (where typically $z$ is much smaller than $\ell$). At the same time we achieve $O(\lg \ell) = O(\lg n)$ communication complexity and thus match or improve the $O(\lg n)$ uncompressed bound. Note that the number of rounds is both compressed and output sensitive and the communication is output sensitive.

As far as we know, this is the first result studying the communication complexity problems in
 LZ77 compressed strings. A previous result by Bar-Yossef et al.~\cite{Bar-YossefJKK04} gives some impossibility results on compressing the text for (approximate) string matching in the sketching model, where a sketching algorithm can be seen as a public-coin one-way communication complexity protocol. Here we exploit the fact that the common prefixes have the same parsing into phrases up to a certain point, and that the ``mismatching'' phrase has a back pointer to the portion of the text represented by the previous phrases: Alice and Bob can thus identify the mismatching symbol inside that phrase without further communication (see the ``techniques'' paragraph). 

We extend the result stated in Theorem~\ref{thm:lcp} so as to compute longest common prefixes when Bob holds a set of $k$ strings $B_1, \ldots, B_k$, and the goal is to compute the maximal longest common prefix between $A$ and any of the strings $B_1, \ldots, B_k$. This problem, denoted $\textsc{Lcp}^k$, naturally captures the distributed scenario, where clients need to search for query strings in a text data base stored at a server. To efficiently handle many queries we want to reduce both communication and rounds for each search. If we again view the strings as integers this is the \emph{predecessor problem}. We generalize Theorem~\ref{thm:lcp} to this scenario. 

\begin{restatable}{theorem}{lcpk}
    \label{thm:lcpk}
    The \textsc{Lcp}$^k$ problem has a randomized public-coin $O(\lg z)$ round communication protocol with $O(\lg z \lg k + \lg \ell)$ communication complexity, where $\ell$
    is the maximal common prefix between $A$ and any one of $B_1, \ldots, B_k$, and $z$ is the number of phrases in the LZ77 parse of this prefix.
\end{restatable}
Compared to Theorem~\ref{thm:lcp} we obtain the same number of rounds and only increase the total communication by an additive $O(\lg z \lg k)$ term. As $z \leq \ell$ the total communication increases by at most a factor $\lg k$.

The mentioned results hold only for LZ77 parses without self-references (see Sec. \ref{sub:LZ}).
We also show how to handle self-referential LZ77 parses and obtain the following bounds, where we add either extra $O(\lg \lg \ell)$ rounds or extra $O(\lg\lg\lg |A|)$ communication.

\begin{restatable}{theorem}{lcpself}
    \label{cor:lcp}
    The \textsc{Lcp} problem has an randomized public-coin protocol with
    \begin{enumerate} 
        \item $O(\lg z + \lg \lg \ell)$ rounds and $O(\lg \ell)$ communication complexity,
        \item $O(\lg z)$ rounds and $O(\lg \ell + \lg \lg \lg |A|)$ communication complexity
    \end{enumerate}
    where $\ell$ is the length of the longest common prefix of $A$ and $B$, and $z$ is the number 
    of phrases in the \emph{self-referential} LZ77 parse of this prefix.
\end{restatable}
\begin{restatable}{theorem}{lcpkself}
    \label{cor:lcpkself}
    The \textsc{Lcp}$^k$ problem has a randomized public-coin protocol with
    \begin{enumerate} 
        \item $O(\lg z + \lg \lg \ell)$ rounds and $O(\lg z \lg k + \lg \ell)$ communication complexity,
        \item $O(\lg z)$ rounds and $O(\lg z \lg k + \lg \ell + \lg \lg \lg |A|)$ communication complexity
    \end{enumerate}
    where $\ell$ is the length of the maximal common prefix between $A$ and any one of $B_1, \ldots, B_k$, and $z$ is the number of phrases in the \emph{self-referential} LZ77 parse of this prefix.
\end{restatable}

Turning again to LZ77 parses without self-references we also show the following trade-offs between rounds and communication.

\begin{restatable}{theorem}{multisearch}
  \label{thm:multisearch}
    For any constant $\epsilon > 0$ the \textsc{Lcp} problem has a randomized public-coin protocol with
		\begin{enumerate} 
                    \item $O(1)$ rounds and $O(z_A^\epsilon)$ total communication	
                    where $z_A$ is the number of phrases in the LZ77 parse of $A$
	            \label{it:multialice}
		\item $O(\lg \lg \ell)$ rounds and $O(z^\epsilon)$ total communication 
                    where $z$ is the number of phrases in the LZ77 parse of the longest common prefix between $A$ and $B$ \label{it:multil}
	\end{enumerate}	
\end{restatable}

Using the standard transformation technique by Newman~\cite{Newman1991} all of the above results can be converted into private-coin results for bounded length strings: If the sum of the lengths of the strings is $\le n$, then, Newman's construction adds an $O(\lg n)$ term in communication complexity, and only gives rise to $1$ additional round. 

\subsubsection{Techniques.}
\label{sub:techniques-outline}
\begin{figure}[h]
	\centering
	\begin{tikzpicture}[font=\ttfamily,
	array/.style={matrix of nodes,nodes={draw, minimum size=3mm, fill=green!30},column sep=-\pgflinewidth, row sep=0.5mm, nodes in empty cells,
		row 1/.style={nodes={draw, fill=none, minimum size=3mm}},
		row 1 column 1/.style={nodes={draw}}}]
	\node[](A) at (-.5,0) {$A$};
	\node[](B) at (-.5,-.6) {$B$};
	\matrix[array] (array) at (2,0) { & 
		\hspace{.8em} & \hspace{1.5em}  &  & \hspace{2em} & \hspace{3em}  &   \\
	};
	\matrix[array] (arrayb) at (2.03,-.6) { &
		\hspace{.8em} & \hspace{1.5em}  &  & \hspace{1em} & \hspace{2.5em}  & &\hspace{1em} \\};
	\draw[<->] (arrayb-1-5.north) -- ([yshift=3mm]arrayb-1-5.north);
	\draw[|-|]([yshift=-3mm]arrayb-1-1.south west) -- node[below] {$1\ldots\ldots\ldots\ldots \ell$} ([yshift=-3mm]arrayb-1-5.south);
	\end{tikzpicture}
\caption{If the longest common prefix $L$ of $A$ and $B$ has $z$ phrases, then the first $z-1$ phrases of $A$, $B$, and $L$ are identical.\label{fig:common}}
\end{figure}
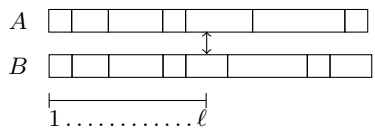

Our results rely on the following key idea. First, we want to perform a binary search over the LZ77-parses of the strings, to find the first phrase where Alice and Bob disagree. Then, the longest common prefix must end somewhere in the next phrase (see Figure~\ref{fig:common}). So Alice needs only to send the offset and length of her next phrase, and Bob can determine the longest common prefix with his string or strings (as proven in Lemma~\ref{lem:lz}).

To implement the idea efficiently, we use standard techniques that allow Alice and Bob to check if a specific prefix of their strings match using $O(1)$ communication, with only constant probability of error (we call this the \textsc{Equality} problem). Similarly, if Bob holds $k$ strings, they can check whether any of the $k$ strings matches Alice's string with only $O(\log k)$ communication, with constant error probability (we call this the \textsc{Membership} problem). This leads to following $O(\log z)$ round communication protocol. 
\begin{enumerate}
	\item Alice and Bob do an exponential search, comparing the first, two first, four first, etc, phrases of their strings using \textsc{Equality} or \textsc{Membership}, until they find a mismatch.
	\item Alice and Bob do a binary search on the last interval of phrases from Step 1, again, using \textsc{Equality} or \textsc{Membership}, until they find their longest common prefix up to a phrase border. 
	\item Alice sends the offset and length of her next phrase, and Bob uses this to determine the longest common prefix. \label{it:last}
\end{enumerate}

To efficiently cope with errors in each step (which can potentially accumulate), we show how to extend techniques for \emph{noisy binary search}~\cite{Feige1994} to an exponential search. Our new \emph{noisy exponential search} only increase the number of rounds by a constant factor.



\subsubsection{Paper outline.}
In Section~\ref{sec:defs}, we review protocols for \textsc{Equality} and \textsc{Membership}. Section~\ref{sec:defs} also contains a formal definition of the LZ77-parse of a string.  
In Section~\ref{sec:noisy}, we recall efficient techniques to handle errors using noisy binary search, and extend them to exponential search.
In Section~\ref{sec:lcp} we go on to prove Theorem~\ref{thm:lcp} and Theorem~\ref{thm:lcpk}. In Section~\ref{sec:selfing}, we show how to extend our results to self-referencing LZ77 (Theorems~\ref{cor:lcp} and~\ref{cor:lcpkself}).
Finally, in section~\ref{sec:mary}, we give the constant-round and near-constant round protocols  promised in Theorem~\ref{thm:multisearch}.


\section{Definition and Preliminaries}\label{sec:defs}

A string $S$ of length $n = |S|$ is a sequence of $n$ symbols $S[1]\cdots S[n]$ drawn from an alphabet $\Sigma$.
The sequence $S[i,j]$ is the \textit{substring} of $S$ given by $S[i]\cdots S[j]$ and, if $i = 1$,
this substring is a \textit{prefix} of $S$. Strings can be concatenated, i.e. $S = S[1, k]S[k+1, n]$.
Let $\textsc{Lcp}(A, B)$ denote the length of the longest common prefix between strings $A$ and $B$. Also, denote by $[u]$ the set of integers $\{1,2, \ldots, u\}$.

\subsubsection{Communication Complexity Primitives.}
We consider the public-coin and private-coin randomized communication complexity models. In the public-coin model the parties share an infinite string of independent unbiased coin tosses and the parties are otherwise deterministic. The requirement is that for every pair of inputs the output is correct with probability at least $1 - \epsilon$ for some specified $1/2 > \epsilon > 0$, where the probability is on the shared random string. We note that any constant probability of success can be amplified to an arbitrarily small constant at the cost of a constant factor overhead in communication. In the private-coin model, the parties do not share a random string, but are instead allowed to be randomized using private randomness. Newman~\cite{Newman1991} showed that any result in the public-coin model can be transformed into private-coin model result at the cost of an additive $O(\log \log T)$ bits of communication, where $T$ is the number of different inputs to the players. In our results this leads to an $O(\log n)$ additive overhead, if we restrict our input to bounded length strings where the sum of the lengths of the strings is $\le n$.

In the \textsc{Membership} problem, Alice holds a string $A$ of length $|A| \le n$, and Bob holds a set $\mathcal{B}$ of $k$ strings. The goal is to determine whether $A\in \mathcal{B}$ (we assume that $n$ and $k$ are known to both parties)~\cite{RaoYehudayoff18}.

\begin{lemma}
	\label{lem:membership}
	The \textsc{Membership} problem has a public-coin randomized $1$-round communication protocol with $m$ communication complexity and error probability $k2^{-m}$,
        for any integer $m > 0$.
\end{lemma}


\noindent\textit{Proof sketch.}
	Let $F : \{0, 1\}^n \rightarrow \{0,1\}^m$ be a random linear function over $GF(2)$ where the coefficients of $F$ are read from the shared random source (public coin).
        Alice applies $F$ to  $A$ and sends the resulting $m$ bits to Bob, i.e., she computes the product between a random $m \times n$ matrix and her string as a vector. 
	Bob applies the same function to each of his strings, i.e., he computes the product between the same random matrix and each of his strings.
        If one of these products is the same as the one he received from Alice he sends a ``1'' to Alice indicating a match. 
	This protocol has no false-negatives and by union bound the
        probability of a false-positive is at most $k2^{-m}$.
        For further details see e.g. \cite{Buhrman,MILTERSEN199837}.
\qed
\medskip
%
%
    

In the \textsc{Equality} problem, Alice holds a string $A$ of size $|A|\le n$, and Bob holds a string $B$. The goal is to determine whether $A = B$ (we assume that $n$ is known to both parties). Lemma~\ref{lem:membership} implies the following corollary. 

\begin{corollary}
	\label{cor:equality}
	The \textsc{Equality} problem has a public-coin randomized $1$-round communication protocol with $m$ communication complexity and error probability~$2^{-m}$.
\end{corollary}

\subsubsection{Lempel-Ziv Compression}\label{sub:LZ}
The \textit{LZ77 parse} \cite{Ziv1977} of a string $S$ of length $n$ divides
$S$ into $z$ substrings $f_1f_2\ldots f_z$, called  \textit{phrases}, in a greedy left-to-right order.
The $i^{th}$ phrase $f_i$ starting at position $u_i$ is the longest substring having a least one occurrence starting 
to the left of $u_i$ plus the following symbol. To compress $S$, we represent each phrase as a tuple $(s_i, l_i, \alpha_i) \in ([n] \times [n] \times \Sigma)$,
such that $s_i$ is the position of the previous occurrence, $l_i$ is its length, and $\alpha_i$ is the symbol at position $u_i+l_i$.
It follows that $s_1 = l_1 = 0, u_1 = 1, \alpha_1 = S[1]$ and we define $e_i=u_i + l_i$ for $i \in z$ .
That is, the $i^{th}$ \emph{phrase} of $S$ ends at position $e_i$.
We call the positions $e_1, \ldots, e_z$ the \emph{borders} of $S$
and the substring $S[s_i, s_i + l_i]$ is the \textit{source} source of the $i^{th}$ phrase $f_i = S[u_i, u_i + l_i]$.

When a phrase is allowed to overlap with its source, the parse is \textit{self-referential}.
A more restricted version does not allow self-references and thus require that $s_i + \l_i \leq u_i$ for $i \in [z]$.
We consider LZ77 parse without self-references unless explecitly stated.
An LZ77 parse of $S$ can be found greedily in $O(n)$ time from the suffix tree of $S$. It is easy to see that $z = \Omega(\lg n)$ if self-references are not allowed, while $z = \Omega(1)$ for self-referential parses.

\section{Noisy Search}\label{sec:noisy}
The \textit{noisy binary search} problem is to find an element $x_t$ among a sequence of elements $x_1, \ldots x_n$ where $x_i \leq x_{i+1}$
using only comparisons in a binary search. Each comparison may fail with a constant probability less than $1/2$ and faults are independent.
%
%
\begin{lemma}[Feige et al.~\cite{Feige1994} Theorem 3.2]
	\label{lem:noisy}
	For every constant $Q < 1/2$, we can solve the noisy binary search problem on $n$ elements with probability at least $1-Q$ in $O(\lg (n/Q))$ steps. 
\end{lemma}
%
%

We now show how to generalize the algorithm by Feige at al. to solve the \textit{noisy exponential search} problem.
That is, given a sequence $x_1, x_2, \ldots$ where $x_i \leq x_{i+1}$ and an element $x_\ell$ find an element $x_r$
such that $\ell \leq r \leq 2\ell$ using exponential search.

\begin{lemma}
	\label{lem:exponential}
	For every constant $Q < 1/2$, we can solve the noisy exponential search problem searching for $x_\ell$ with probability at least $1-Q$ in $O(\lg (\ell/Q))$ steps. 
\end{lemma}
\begin{proof}
    In case of no errors we can find $x_r$ on $O(\lg \ell)$ steps comparing $x_\ell $ and $x_i$ for $i = 1, 2, 4, 8 \ldots$ until $x_i \geq x_\ell$. At this point we have $\ell\leq i \leq 2\ell$.


    Consider the decision tree given by this algorithm. 
    This tree is simply a path $v_0, v_1, v_2, \ldots$ and
    when reaching vertex $v_i$ the algorithm compares elements $x_l$ and $x_{2^i}$. 
    In order to handle failing comparisons we tranform this tree by adding a path with length $l_i$ (to be specified later) as a child of vertex $v_i$.
    Denote such a path with $p_i$.
    The search now performs a walk in this tree starting in the the root and progresses as follows: 
    Reaching vertex $v_i$ we first check if $x_l \geq x_{2^{i-1}}$.
    If not, this reveals an earlier faulty comparison and we backtrack by moving to the parent.
    Otherwise, we check if $x_l \geq x_{2^i}$. If so we move to
    vertex $v_{i+1}$. Otherwise, we move to the first vertex on the path $p_i$.
    Reaching a vertex $u$ on a path $p_i$ we test if $x_l \geq x_{2^{i-1}}$
    and if $x_l < x_{2^i}$.
    If both tests are positive, we move to the only child of $u$.
    Otherwise, this reveals an earlier faulty comparison and we backtrack
    by moving to the parent of $u$. 

    The search can be modeled as a Markov process.
    Assume that $\lceil \lg \ell \rceil = j$ and thus $j = O(\lg \ell)$ and direct all edges towards the leaf $u$ on the path $p_j$.
    For every vertex, exactly one adjacent edge is
    directed away from $u$ and the remaining edges are directed towards $u$.
    The transition probability along an outgoing edge of a vertex is at greater than $1/2$ and 
    the transition probability along the remaining edges is less than $1/2$.
    Let $b$ be the number of backward transitions and $f$ the number of forward transitions.
    We need to show that $f - b \geq j + l_j$ with probability at least
    $1 - Q$ for $Q < 1/2$ implying that the search terminates in the leaf $u$.
    Setting $l_i = ic_1$ this follows after $c_2(\lg(2^j/Q)) = O(\lg(\ell/Q))$ rounds from Chernoff's bound \cite{chernoff}
    with suitable chosen constants $c_1$ and $c_2$.
    \qed
\end{proof}

\section{Communication Protocol for {\sc $\LCP$}}\label{sec:lcp}
We now present our protocol for the $\LCP$ problem without self-references. We consider the case with self-references in the next section. First, we give an efficient uncompressed output sensitive protocol that works for an arbitrary alphabet (Lemma~\ref{lem:lcp-no-compress}). Secondly, we show how to encode LZ77 strings as strings from a small alphabet (Lemma~\ref{lem:phrase}) which allows us to efficiently determine the first phrase where Alice and Bob disagree. Thirdly, we show that given this phrase Alice and Bob can directly solve $\LCP$ (Lemma~\ref{lem:lz}). Combining these results leads to Theorem~\ref{thm:lcp}. Finally, we  generalize the results to the $\LCP^k$ case.

First we show how to solve the $\LCP$ problem with output-sensitive complexity for both the number of rounds and the amount of bits of communication.
\begin{lemma}
    \label{lem:lcp-no-compress}
    Let $A$ and $B$ be strings over an alphabet $\Sigma$ known to the parties.
    The \textsc{Lcp} problem has a pulic-coin randomized $O(\lg \ell)$-round communication protocol with $O(\lg \ell)$ communication complexity, where $\ell$ is the length of the longest common prefix between $A$ and $B$.
\end{lemma}

\begin{proof}
    Alice and Bob compare prefixes of exponentially increasing length using equality, and stop
    after the first mismatch. Let $t$ be the length of the prefixes that do not match and observe that $t \leq 2\ell$.
    They now do a binary search on the interval $[0,  t]$, using equality to decide if the left or right
    end of the interval should be updated to the midpoint in each iteration. 
    The parties use Corollary~\ref{cor:equality} with $m = 2$,
    and new random bits from the shared random source for every equality check.
    Thus, the probability of a false-positive is at most $1/4$, and the faults are independent. Using Lemma~\ref{lem:exponential} and Lemma~\ref{lem:noisy} we get that we can solve the problem in $O(\lg (\ell/Q))$  rounds of communication
    with probability at least $1 - Q$ for any constant $Q < 1/2$.
%
\qed
\end{proof}

Note that the size of the alphabet $\Sigma$ does not affect the complexity of this protocol.
Alice and Bob do however need to agree on how many bits to use per symbol in order
to use the same number of random bits for the equality checks. Because $\Sigma$ is known to the parties, they sort the alphabet and use $\lg |\Sigma|$ bits per symbol.

We move on to consider how to handle LZ77 compressed strings. Recall that the $i^{th}$ phrase in the LZ77 parse of a string $S$ is represented as a tuple $(s_i, l_i, \alpha_i)$ consisting of the source $s_i$, the length $l_i$ of the source, and a symbol $\alpha_i \in \Sigma$. Observe that the LZ77 parse can be seen as a string where each tuple describing a phrase corresponds to a symbol in this string. Because we consider LZ77 without self-references a phrase is never longer than sum of the lengths of the previous phrases and we can thus bound the number of bits required to write a phrase.

\begin{lemma}
    \label{lem:phrase}
    Let $Z_i = (s_1, l_1, \alpha_1), \ldots, (s_i, l_i, \alpha_i)$ be the first $i$ elements in the LZ77
    parse of a string $S$. 
    Then, $s_i$ and $l_i$ can be written in binary with $i$ bits. 
\end{lemma}

\begin{proof} 
Recall that $e_j$ is the position in $S$ of the last symbol in the $j^{th}$ phrase. Since  we have no self-references $s_i$ and $l_i$ are both no larger than $e_{i-1}$ they can be written with $\lg e_{i-1}$ bits. 
 By definition $u_j = e_{j-1}+1$. Therefore, $e_j = u_j + l_j = e_{j-1}+ 1+ l_j \leq 2e_{j-1} + 1$, and it follows that $e_{i-1} \leq 2 e_{i-2} + 1\leq \cdots \leq 2^{i} -1$ since $e_1 = 1$. 
\qed
\end{proof}

We show that $\ell = \LCP(A,B)$ can be determined from $\LCP(Z_A, Z_B)$ with only one round and $O(\lg \ell)$ communication,
where $Z_A$ and $Z_B$ are the respective LZ77 parses of $A$ and $B$.

While a LZ77 parse of a string is not necessarily unique, in this case, we can assume that the parties
as part of the protocol agree deterministically upon their same decisions on LZ77-compression algorithm (e.g. taking always the leftmost source when there are multiple possibilities). This ensures that we obtain the same parsing for equal strings, independently and without any communication.

\begin{lemma}
    \label{lem:lz}
    Let $A$ and $B$ be strings and let $Z_A$ and $Z_B$ be their respective LZ77 parses.
    If Alice knows $A$ and Bob knows $B$ and the length of the longest common prefix $\LCP(Z_A, Z_B)$, then they can determine 
    the length $\ell= \LCP(A, B)$ of the longest common prefix of $A$ and $B$ in $O(1)$ round and $O(\lg \ell)$ communication.
\end{lemma}

\begin{proof}
    First, $Z_A$ and Z$_B$ themselves can be seen as strings over the special alphabet $\Sigma' \equiv ([n] \times [n] \times \Sigma)$ of tuples. 
    Letting $z = \LCP(Z_A, Z_B)$, these LZ77 parses of $A$ and $B$ are identical up until but no longer than their $z^{th}$ tuple.   Now, let $\ell = \LCP(A, B)$.
    Let $a_i$ and $b_i$ denote the $i^{th}$ phrase border in the LZ77 parse of $A$ and $B$ respectively.
    Observe that $A[1, a_z] = B[1, b_z]$ but $A[1, a_{z+1}] \neq B[1, b_{z+1}]$ because of how we choose $z$ and, thus, $a_z = b_z \leq \ell < a_{z+1}, b_{z+1}$.
    Let $s_{z+1}, l_{z+1}$ be the source and length of the $(z+1)^{th}$ phrase in $Z_A$.
    Alice sends $s_{z+1}, l_{z+1}$ to Bob in one round with $O(\lg a_z) = O(\lg \ell)$ bits of communication since $s_{z+1}, l_{z+1} \leq a_z$.
    At this point, it is crucial to observe that Bob can recover $A[1, a_{z+1}]$ by definition of LZ77 parsing:
    he deduces that $A[1, a_{z+1}] = B[1, b_{z}]B[s_{z+1}, s_{z+1} + l_{z+1}]$,
    from which he can compute $\LCP(A[1,a_{z+1}], B[1,b_{z+1}])=\LCP(A, B)$.
\qed
\end{proof}

We can now combine Lemmas~\ref{lem:lcp-no-compress},~\ref{lem:phrase}, and~\ref{lem:lz} to prove Theorem~\ref{thm:lcp}.
%
%
    Alice and Bob construct the LZ77 parse of their respective strings and interprets the parse 
    as a string. Denote these strings by $Z_A$ and $Z_B$.
    They first use Lemma~\ref{lem:lcp-no-compress} to determine $\LCP(Z_A, Z_B)$,
    where the parties decide to use $2i + \lg |\Sigma|$ random bits for the equality check of the $i^{th}$ symbols (from $\Sigma'$), which suffices by Lemma~\ref{lem:phrase}.
    Then they apply Lemma~\ref{lem:lz} to determine $\LCP(A, B)$. In conclusion this proves Theorem~\ref{thm:lcp}.


\subsection{The {\sc $\textsc{Lcp}^k$} case}
In this section we generalize the result on LCP to the case where Bob holds multiple strings. Here, Alice knows a string $A$ and Bob knows strings $B_1,\ldots,B_k$,
where all strings are drawn from an alphabet $\Sigma$ known to the parties. 

The main idea is substitute the equality-tests by membership queries. 
We first generalize Lemma~\ref{lem:lcp-no-compress} to the $\LCP^k$-case. 
%

\begin{lemma}\label{lem:lcpk}
	The \textsc{Lcp}$^k$-problem has a randomized public-coin $O(\lg \ell)$-round communication protocol with $O(\lg \ell\lg k)$
	communication complexity, where $\ell$ is the length of the maximal longest common prefix between $A$ and any $B_i$.
\end{lemma}

\begin{proof}Along the same lines as the proof of Lemma~\ref{lem:lcp-no-compress}, Alice and Bob perform membership-queries on exponentially increasing prefixes, and then, perform membership-queries to guide a binary search. 
They use Lemma~\ref{lem:membership} with $m=2\lg k$, and exploit shared randomness as in the previous case.
Again, the probability of a false positive is $\le 1/4$, and the faults are independent.  Thus  Lemma~\ref{lem:exponential} and Lemma~\ref{lem:noisy} gives us a $O(\lg \ell)$ round communication protocol with total error probability $1 - Q$ for any constant choice of $Q < 1/2$.

Since there are $O(\lg \ell)$ rounds in which we spend $O(\lg k)$ communication, the total communication becomes $O(\lg \ell \lg k)$.
\qed
\end{proof}

We go on to show that the maximal $\LCP(A,B_i)$ can be determined from solving $\LCP^k$ on $Z_A$ and $\{Z_{B_1},\ldots , Z_{B_k}\}$ with only one additional round and $O(\lg n)$ communication.

\begin{lemma}
	\label{lem:lzk}
	Let $Z_A,Z_{B_1},\ldots ,Z_{B_k}$ be the LZ77 parses of the strings $A, B_1,\ldots,B_k$.
	If Alice knows $A$, and Bob knows $B_1,\ldots, B_k$ and the length of the maximal longest common prefix between $Z_A$ and any $Z_{B_i}$,
	 they can find $\max_i \LCP(A, B_i)$ in $O(1)$ round
	and $O(\lg n)$ communication.
\end{lemma}
\begin{proof}
	In this case, Bob holds a set, $\mathcal{B}'$, of at least one string that matches Alice's first $z$ phrases,
	and no strings that match Alice's first $z+1$ phrases. Thus, if Alice sends the offset and length of her next phrase, he may determine $\LCP(A,B_i)$ for all strings $B_i\in\mathcal{B}'$. Since the maximal
	$\LCP$ among $B_i\in \mathcal{B}'$ is indeed the maximal over all $B_i \in \mathcal{B}$, we are done.
\qed
\end{proof}
Combining Lemma~\ref{lem:lcpk} and Lemma~\ref{lem:lzk} we get Theorem~\ref{thm:lcpk}. 

\section{Self-referencing LZ77}\label{sec:selfing}

We now consider how to handle LZ77 parses with self-references.
The main hurdle is that Lemma~\ref{lem:phrase} does not apply in this case
as there is no bound on the phrase length except the length of the string.
This becomes a problem when the parties need to agree on the number of bits
to use per symbol when computing $\LCP$ of $Z_A$ and $Z_B$, but also
when Alice needs to send Bob the source and length of a phrase in order for him to decide $\LCP(A,B)$.

First we show how Alice and Bob can find a bound on the number of random bits to use
per symbol when computing $\LCP(Z_A,Z_B)$.

\begin{lemma}\label{lem:bounding}
    Bob and Alice can find an 
    upper bound  $\ell' $on the length of the longest common prefix between $A$ and $B$ where 
    \begin{enumerate}
               \item $\ell' \leq \ell^2$ using $O(\lg \lg \ell)$ rounds and $O(\lg \lg \ell)$ total communication\label{it:rounds}
                \item $\ell' \leq |A|^2$ using $O(1)$ round and $O(\lg \lg \lg |A|)$ total communication.\label{it:alicesends}
    \end{enumerate}

\end{lemma}
\begin{proof}
    Part~(\ref{it:rounds}): Alice and Bob do a double exponential search for $\ell$ and find a number $\ell\leq \ell' \leq \ell^2$ using equality checks on prefixes of their uncompressed strings in $O(\lg \lg \ell)$ rounds. Again, at the cost of only a constant factor, we apply Lemma~\ref{lem:exponential} to deal with the probability of false positives.

    Part~(\ref{it:alicesends}): Alice sends the minimal $i$ such that $|A| \leq 2^{2^i}$ thus $i = \lceil \lg \lg |A| \rceil$  can be written in
    $O(\lg \lg \lg |A|)$ bits. 
    Alice and Bob can now use $n=2^{2^i}$ as an upper bound for $\ell$, since $\ell \le |A| \le 2^{2^i} < |A|^2$.
\qed
\end{proof}

Assume that Alice and Bob find a bound $\ell'$ using one of those techniques, then they can safely truncate their strings 
to length $\ell'$. Now they know that every symbol in $Z_A$ and $Z_B$ can be written with $O(\lg \ell' + \lg |\Sigma|)$ bits,
and thus, they agree on the number of random bits to use per symbol when doing equality (membership) tests. Using Lemma~\ref{lem:lcp-no-compress} they can now find the length of the longest common prefix between $Z_A$ and $Z_B$ in $O(\lg \ell)$ rounds with $O(\lg \ell)$ communication. 

We now show how to generalize Lemma~\ref{lem:lz} to the case of self-referential parses. 

\begin{lemma}
    \label{cor:lzself}
    Let $A$ and $B$ be strings and let $Z_A$ and $Z_B$ be their respective \emph{self-referential} LZ77 parses.
    If Alice knows $A$ and Bob knows $B$ and the length of the longest common prefix between $Z_A$ and $Z_B$, then they can determine 
    the length $\ell$ of the longest common prefix of $A$ and $B$ in
    \begin{enumerate}
        \item $O(1 + \lg \lg \ell)$ rounds and $O(\lg \ell)$ communication
        \item $O(1)$ rounds and $O(\lg \ell + \lg \lg \lg |A|)$ communication
    \end{enumerate}
\end{lemma}

\begin{proof}
    Let $s_i, e_i$ and $l_i$ be the respective source, border and length of the $i^{th}$ phrase in $Z_A$.
    The proof is the same as in Lemma~\ref{lem:lz} except that the length $l_{z+1}$ of the $(z+1)^{th}$ phrase in $Z_A$ that Alice sends to Bob 
    is no longer bounded by~$\ell$.
        
    There are two cases. If $l_{z+1} \leq 2e_z$, then $l_{z+1} \leq 2\ell$,  and  Alice can send $l_{z+1}$ to Bob in one round and $O(\lg \ell)$ bits and we are done.
    
    If  $l_{z+1} > 2e_z$ then the source of the $(z+1)^{th}$ phrase must overlap with the phrase itself
    and thus the phrase is periodic with period length at most $e_z$ and has at least $2$ full repetitions of its period.
    Alice sends  the starting position of the source of the phrase $s_{i+1}$ along with a message indicating that we
    are in this case to Bob in $O(\lg \ell)$ bits.
    Now Bob can check if they agree on next $2e_z$ symbols. If this is not the case, he has also determined $\LCP(A, B)$
    and we are done. Otherwise, they agree on the next $2e_z$ symbols and therefore
    $(z+1)^{th}$ phrases of both A and B are periodic with the same period. 
    What remains is to determine which phrase that is shorter. Let $l_a$ and $l_b$ denote the lengths of respectively Alice's and Bob's
    next phrase.
     Then
    (1) follows from Alice and Bob first computing a number $\ell' \leq \ell^2$ using a double exponential search and equality checks in $O(\lg \lg \ell)$ rounds
    and total communication. Clearly either $l_a$ or $l_b$ must be shorter than $\ell'$ and the party with the shortest phrase
    sends its length to the other party in $O(\lg \ell)$ bits and both can then determine $\LCP(A, B)$.
    To get the result in (2) Alice sends the smallest integer $i$ such that $l_a \leq 2^{2^i}$  in a single round and $O(\lg \lg \lg |A|)$ bits of communication.
    Bob then observes that if $l_b \leq 2^{2^{i-1}}$, then $l_b = \ell$ and he sends $\ell$ to Alice using $O(\lg \ell)$ bits.
    If $l_b > 2^{2^i}$ then $l_a = \ell$ and he informs Alice to send him $l_a$ in $O(\lg \ell)$ bits. Finally, if 
    $2^{2^i-1} < l_b$ and $l_a \leq 2^{2^i} \leq \ell^2$ he sends $l_b$ to Alice using $O(\lg \ell)$ bits.
\qed
\end{proof}

Theorem~\ref{cor:lcp} now follows from Lemmas~\ref{lem:lcp-no-compress},~\ref{lem:bounding}, and~\ref{cor:lzself}.
%

\subsection{$\LCP^k$ in the self-referential case.}

Finally, we may generalize Theorem~\ref{thm:lcpk} to the self-referential case.
Substituting equality with membership, we may directly translate Lemma~\ref{lem:bounding}:
\begin{lemma}\label{lem:boundingk}
	Bob and Alice can find an upper bound on the length $\ell'$ of the maximal longest common prefix between $A$ and $B_1,\ldots,B_k$ where
	\begin{enumerate}
		\item $\ell' \leq \ell^2$ using $O(\lg \lg \ell)$ rounds and $O(\lg \lg \ell \log k)$ total communication\label{it:roundsk}
		\item $\ell' \leq |A|^2$ using  $O(1)$ round and $O(\lg \lg \lg |A|)$ total communication.\label{it:alicesendsk}
	\end{enumerate}
\end{lemma}

Using the lemma above, we can generalize Corollary~\ref{cor:lzself} to the LCP$^k$-case.
\begin{lemma}
    \label{cor:lzkself}
    Let $A$ and $B_1,\ldots,B_k$ be strings, and let $Z_A$ and $Z_{B_i}$ be their respective \emph{self-referential} LZ77 parses.
    If Alice knows $A$ and Bob knows $B_1,\ldots,B_k$ and Bob knows the length of the maximal longest common prefix between $Z_A$ and any $Z_{B_i}$, then they can determine $\ell$ in
    %
    \begin{enumerate}
    	\item $O(1 + \lg \lg \ell)$ rounds and $O(\lg \ell\lg k)$ communication
    	\item $O(1)$ rounds and $O(\lg \ell\lg k + \lg \lg \lg |A|)$ communication
    \end{enumerate}
\end{lemma}

\noindent\textit{Proof tweak.} Alice and Bob have already found a common prefix of size $e_z$ -- question is whether a longer common prefix exists. As before, if Alice's next phrase is shorter than $2e_z$, she may send it. Otherwise, she sends the offset, and indicates we are in this case. Now, Bob can check if \emph{any} of his strings agree with Alice's on the next $2e_z$ symbols. If none do, we are done. If several do, he forgets all but the one with the longest $(z+1)$'st phrase, and continue as in the proof of Corollary~\ref{cor:lzself}.\qed

Theorem~\ref{cor:lcpkself} now  follows from the combination of Lemmas~\ref{lem:boundingk} and~\ref{cor:lzkself}.

\section{Obtaining a trade-off via $D$-ary search.}\label{sec:mary}
We show that the technique of Buhrman et al.~\cite{Buhrman}, to compute \textsc{Lcp} of two strings of length $n$
in $O(1)$ rounds and $O(n^\epsilon)$ communication, can be used to obtain a compressed communication complexity. 
Note that we again consider LZ77 compression without self-references.
We first show the following generalization of Lemma~\ref{lem:lcp-no-compress}.

\begin{lemma}
    \label{lem:b-ary}
    Let $A$ and $B$ be strings over an alphabet $\Sigma$ known to the parties.
    The \textsc{Lcp} problem has a public-coin randomized communication protocol
    with 
    \begin{enumerate} 
        \item $O(1)$ rounds and $O(|A|^\epsilon)$ communication \label{it:alicem}
        \item $O(\lg \lg \ell)$ rounds and $O(\ell^\epsilon)$ communication \label{it:m}
    \end{enumerate} 
    where $\ell$ is the length of the longest common prefix between $A$ and $B$,
    and $\epsilon > 0$ is any arbitrarily small constant.
\end{lemma}
\begin{proof}
    Assume the parties agree on some parameter $C$ and have previous knowledge of some constant $\epsilon'$ with $0 < \epsilon' < \epsilon$ (i.e. $\epsilon'$ and $\epsilon$ are plugged into their protocol). They perform a $D$-ary search in the interval $[-1, C]$ with $D = C^{\epsilon'}$.
	In each round, they split the feasible interval into $D$ chunks, and perform equality tests from Corollary~\ref{cor:equality} with $m=2\lg(D/\epsilon')$
    on the corresponding prefixes. The feasible interval is updated to be the leftmost chunk where the test fails. 
    There are $\lg_{D} C = 1/\epsilon' = O(1)$ rounds.
    The communication per round is $2D\lg(D/\epsilon')$ and the total communication is 
	$1/\epsilon' \cdot 2D\lg(D/\epsilon') = O(C^{\epsilon'} \lg C)$.
    The probability of a false positive for the equality test is $2^{-m}$, and thus, by a union bound
    over $D$ comparisons in each round and $1/\epsilon'$ rounds, the combined probability of failure becomes $1/4$.
	
	\begin{enumerate}
		\item     
                    Alice sends $|A|$ to Bob in $\lg |A| = O(|A|^\epsilon)$ bits and they use $C = |A|$.
		The total communication is then $O(C^{\epsilon'} \lg C)  = O(|A|^\epsilon)$ with $O(1)$ rounds.
	    \item 
	    Alice and Bob use Lemma \ref{lem:bounding} to find an $\ell'$ such that $\ell \leq \ell' \leq \ell^2$ in $O(\lg\lg \ell)$ rounds and communication. They run the $D$-ary search protocol where $\epsilon' < \epsilon/4$,
	    setting $C = \ell'$. 
	    The extra communication is $O(C^{\epsilon'} \lg C) = O(\ell^\epsilon)$. 
\qed 
	\end{enumerate}
\end{proof}

    We can now combine Lemmas~\ref{lem:b-ary}, \ref{lem:phrase}, and \ref{lem:lz} to prove Theorem~\ref{thm:multisearch}.
%
    Alice and Bob construct the LZ77 parse of their respective strings and interpret the parses 
    as a strings, denoted by $Z_A$ and $Z_B$.
    They first use Lemma~\ref{lem:b-ary} to determine $\LCP(Z_A, Z_B)$,
    and then Lemma~\ref{lem:lz} to determine $\LCP(A, B)$.
    The parties 
    use $2i + \lg |\Sigma|$ random bits for the $i^{th}$ symbol, which suffices by Lemma~\ref{lem:phrase}.
    This enables them to apply Lemma~\ref{lem:b-ary} to $Z_A$ and $Z_B$. In conclusion this proves Theorem~\ref{thm:multisearch}.

We note without proof that this trade-off also generalizes to self-referential parses by paying 
an additive extra $O(\lg \lg \lg |A|)$ in communication for Theorem~\ref{thm:multisearch} (\ref{it:multialice})
and an additive $O(\lg \ell)$ communication cost for Theorem~\ref{thm:multisearch} (\ref{it:multil}).
The same goes for \textsc{Lcp}$^k$ where the comminication increases by a factor $O(\lg k)$
simply by increasing $m$ by a factor $\lg k$ and using the techniques already described.

\bibliography{biblio}
\end{document}